%% file: main.tex
\documentclass[conference, 10pt]{IEEEtran}
\IEEEoverridecommandlockouts
\usepackage{cite}
\usepackage{amsmath,amssymb,amsfonts}
\usepackage{algorithmic}
\usepackage{graphicx}
\usepackage{textcomp}
\usepackage{xcolor}
\usepackage{caption}
\usepackage{subfigure}
\usepackage{algorithm}
\usepackage{algorithmic}
\def\BibTeX{{\rm B\kern-.05em{\sc i\kern-.025em b}\kern-.08em
    T\kern-.1667em\lower.7ex\hbox{E}\kern-.125emX}}
\graphicspath{{./}{figures/}}
\begin{document}

\title{Performance Analysis and Improvement on DSRC Application for V2V Communication \\}

\author{\IEEEauthorblockN{Liu Cao$^*$, Hao Yin$^*$, Jie Hu$^\dagger$, Lyutianyang Zhang$^*$}
\IEEEauthorblockA{{$^*$Department of Electrical and Computer Engineering, University of Washington} \\ {$^\dagger$Department of Electrical and Computer Engineering, North Carolina State University} \\
Email: $^*$\{liucao, haoyin, lyutiz\}@uw.edu,  $^\dagger$jhu29@ncsu.edu } 
}

\maketitle

\input{abstract}

\input{introduction}

\input{analytical}

\input{results}

\input{conclusion}

\bibliographystyle{ieeetr}
\bibliography{ref}

\end{document}

%% file: abstract.tex
\begin{abstract}
In this paper, we focus on the performance of vehicle-to-vehicle (V2V) communication adopting the Dedicated Short Range Communication (DSRC) application in periodic broadcast mode. An analytical model is studied and a fixed point method is used to analyze the packet delivery ratio (PDR) and mean delay based on the IEEE 802.11p standard in a fully connected network under the assumption of perfect PHY performance. With the characteristics of V2V communication, we develop the Semi-persistent Contention Density Control (SpCDC) scheme to improve the DSRC performance. We use Monte Carlo simulation to verify the results obtained by the analytical model. The simulation results show that the packet delivery ratio in SpCDC scheme increases more than 10\% compared with IEEE 802.11p in heavy vehicle load scenarios. Meanwhile, the mean reception delay decreases more than 50\%, which provides more reliable road safety.

\end{abstract}

\begin{IEEEkeywords}
V2V, DSRC, IEEE 802.11p, DCF, MAC design
\end{IEEEkeywords}

%% file: introduction.tex
\section{Introduction}
Vehicle-to-vehicle (V2V) communication is a cornerstone of connected vehicles (CVs) which are emerging as an important component of the next generation intelligent transportation systems (ITS)\cite{lu2014connected}. As an effort to deploy CVs, technologies and standards have been actively developed. Dedicated short-range communications (DSRC) has been tested as an enabling technology for V2V and V2I communications\cite{kenney2011dedicated}. DSRC is a high-efficiency wireless communication technology used in the smart transportation system. In V2V communications for CV applications, the most important component is the broadcast of the Basic Safety Messages (BSMs)\cite{kloiber2015random}. The BSMs are single-hop, periodic, and carry safety-related status information of vehicles such as their speed, acceleration, position, and direction. Through the broadcast of BSMs by DSRC, vehicles can be aware of each other’s status, and traffic accidents can be reduced. 

In DSRC, IEEE 802.11 distributed coordination function (DCF) MAC protocol has been adopted by the IEEE 802.11p standard for DSRC applications. The MAC performance of DSRC has been studied in some papers. The authors in \cite{chen2007quantitative} established a quantitative approach to describe the characteristics of DSRC safety communication. The model in \cite{hassan2011performance} provided an analytical model for the MAC protocol of DSRC under aperiodic broadcast mode. In \cite{bazzi2017performance}, the authors also presented the performance of IEEE 802.11p considering both MAC and PHY layers. However, most of them only focused on analyzing the performance where IEEE 802.11p was applied while not emphasizing how to improve the DSRC performance. The performance such as the PDR and packet delay will degrade heavily in high vehicle load scenarios. 

In this paper, we study an analytical model for IEEE 802.11p in periodic broadcast mode to analyze the DSRC performance. Since each vehicle collects the information from other vehicles through the received BSMs from the previous periods, they can obtain a timeline of packet generation from others. Each vehicle can determine its backoff counter based on the historical information rather than randomly choosing a number in the range of a fixed contention window utilized by IEEE 802.11p. Using this characteristics of V2V communication, we develop the SpCDC scheme which shows better performance than IEEE 802.11p especially in heavy vehicle load scenarios.

This paper is organized as follows: Section II provides an analytical model for IEEE 802.11p adopting the DCF for channel access in periodic broadcast mode. Section III develops a new distributed scheme that enhances the DSRC performance. Section IV compares the results of the Monte Carlo simulation with the results obtained by their analytical models. Section V draws the conclusions.

%% file: analytical.tex
\section{ANALYTICAL MODEL FOR IEEE 802.11p}
In this section, we study an analytical model for IEEE 802.11p in periodic broadcast mode. We assume perfect PHY-layer performance to simplify the analysis, i.e., any packet sent within a given radius can be heard perfectly if not interfered by others. Besides, we use a fixed point model to characterize the DSRC  performance for V2V communications. The mechanism of DCF employing entire carrier sense multiple access with collision avoidance (CSMA/CA) procedure. Each vehicle prepared to send a packet first senses the channel for a period, which is known as the distributed inter-frame space (DIFS). If the channel is sensed busy during this period, the access will be deferred and wait for a complete transmission from the other vehicle. A backoff process will initiate after the channel becomes idle again for a DIFS. Before the backoff process, the vehicle needs to choose a random number within a fixed contention window as the initial backoff counter which decrements by one every time. The counter during the backoff process is suspended when a transmission is detected in the channel and will be reactivated after the channel is sensed idle again for a DIFS. When the counter reaches zero, the vehicle sends the packet instantly. Otherwise, if a vehicle senses the channel idle in the first whole DIFS period, it will occupy the channel and send the packet directly. In broadcast mode, the transmitter vehicle doesn't need acknowledgements(ACKs) from other vehicles since gathering the information from all vehicles will lead to a prohibitively high overhead. Thus, there is no re-transmission or increment of the contention window even if a packet collision occurs.

\subsection{Packet delivery ratio}
PDR is defined as the probability that a BSM (we will always refer BSM as packet) from the tagged vehicle is successfully broadcasted to all other vehicles in its transmission range. We define $\rho$ as the probability of a packet staying at the buffer for each vehicle, which can be expressed as 
\begin{equation}
\rho=\lambda \mathbb{E}[S]\label{eq:rho},
\end{equation}
where $\mathbb{E}[S]$ is the average service time for a packet staying at the buffer. $\lambda$ is the packet transmission frequency, which indicates each vehicle regularly generates packet every $1/\lambda$ seconds. Define $p_{b}$ as the probability that the channel is sensed busy when a new packet arrives, which is given by
\begin{equation}
p_{b}=\left ( N_{tr}-1 \right )\lambda T_{tr}\left( 1-\frac{(n_{c}-1)}{n_{c}}p_{c}\right )\label{eq:pb},
\end{equation}
where $N_{tr}$ is the number of vehicles in the network. $n_{c}$ is the average number of collided packets if a collision occurs. $p_c$ is the collision probability which will be introduced later. A fixed transmission delay $T_{tr}$ is
\begin{equation}
T_{tr}=\frac{\mathbb{E}\left [ P \right ]}{R_{d}}+T_{H}+\delta \label{eq:ttr},
\end{equation}
where $\mathbb{E}\left [ P \right ]$  is the mean packet length of payload and $R_{d}$ is the data rate. $T_{H}$ is the duration of transmitting the packet headers including physical layer header and MAC layer header. $\delta$ is the propagation delay, in this paper, $\delta=0$. In periodic broadcast mode, two cases will happen when a packet arrives:
\begin{itemize}
\item Case 1: Vehicle immediately sends the packet without performing a backoff process if the channel is sensed idle for a DIFS period. 
\item Case 2: The packet will performance a backoff process before transmission if the channel is sensed busy. The corresponding probability is given by $p_{b}$.
\end{itemize}
Here we don't consider the case where a previous packet waiting for a long time due to the backoff process will be replaced by a new arriving packet since the packet inter-arrival time is deterministic ($1/\lambda$ seconds) and it is much longer than any possible packet delay. DCF employs a discrete time slot backoff scheme, if a backoff process is involved, the transmission is synchronized to the beginning of a time slot\cite{yin2013performance}. Therefore, packet collision only occurs in the second case in this paper.

\begin{figure}[htbp]
\captionsetup{font={scriptsize}}
\centering
\includegraphics[width=0.4\textwidth]{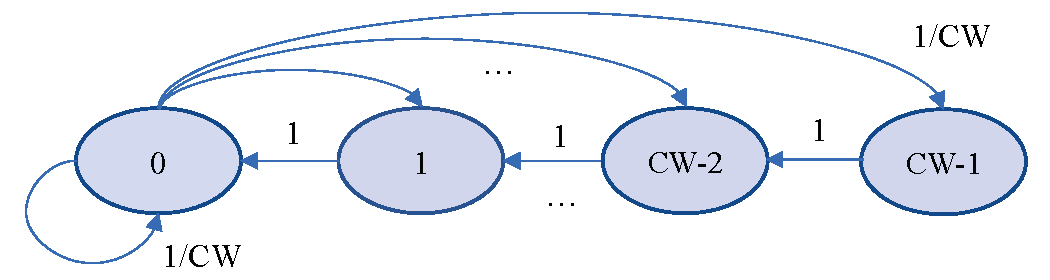}
\caption{Markov Chain for backoff counter}
\label{fig:broad}
\end{figure}

We construct a model to characterize the backoff counter for the IEEE 802.11p broadcast network. The backoff counter, which indicates the counter value of a broadcast vehicle, is a one-dimensional discrete time Markov Chain. The state transition diagram describing the decrements of a backoff counter is shown in Fig.\ref{fig:broad}.
The non-null one-step transition probabilities are
\begin{equation}
\left\{\begin{matrix}
 P(M_{k+1}=m-1|M_{k}=m)=1& m \in [1, CW-1]\\ 
 P(M_{k+1}=m|M_{k}=0)=1/CW& m \in [0, CW-1]
\end{matrix}\right\}\label{eq:pmk},
\end{equation}
where $CW$ is a fixed contention window, and $M_k$ is the value that backoff counter reaches at discrete time $k$. The following relations can be derived according to Eq. (\ref{eq:pmk})
\begin{equation}
\left\{\begin{matrix}
\pi _{m}=\frac{CW-m}{CW}\pi _{0}\\ 
\sum_{m=0}^{CW-1}\pi _{m}=1
\end{matrix}\right\}\label{eq:pi},
\end{equation}
where $m \in \left [ 0, CW-1 \right ]$. $\pi _{m}$ is the probability that the counter reaches $m$ during the backoff process. $\pi _{0}$ is the probability that a vehicle starts to transmit the packet since the counter reaches zero. We can obtain $\pi _{0}$ by solving Eq. (\ref{eq:pi})
\begin{equation}
\pi _{0}=\frac{2}{1+CW}\label{eq}.
\end{equation}

For any vehicle other than the tagged vehicle, the probability of transmitting a packet is $\rho\pi_{0}$ given that the second case happens. A collision occurs when at least one vehicle send packet in the same slot as the tagged vehicle. Thus, the collision probability can be written as
\begin{equation}
p_{c}=p_b\left ( 1-\left ( 1-\rho \pi _{0}  \right ) ^{N_{tr}-1}\right )\label{eq:pc},
\end{equation}
meanwhile PDR is
\begin{equation}
PDR=1-p_{c}\label{eq:pdr}.
\end{equation}

\subsection{Mean delay}
The service time $S$ includes the access delay $T_A$ and the transmission delay $T_{tr}$. The access delay is defined as the interval between the instant the packet reaches the head of the queue and the instant when the packet transmission begins. The end-to-end delay $T_{s}$ experienced by a packet is
\begin{equation}
T_{s}=Q+S=Q+T_{A}+T_{tr} \label{eq:ts},
\end{equation}
where $Q$ and $T_A$ are random variables(r.v.) indicating the queuing delay and the access delay. For periodic broadcast mode, the queuing delay is zero for each packet. The access delay is classified as:
\begin{itemize}
\item For case 1, the access delay is a DIFS period since packet doesn't perform a backoff process.
\item For case 2, the packet needs to wait for ongoing packet transmission and then performs a backoff process.
\end{itemize}
More precisely, the access delay can be summarized as follows:
\begin{equation}
T_{A}=\left\{\begin{matrix}
DIFS &w.p.\quad  1-p_{b} \\ 
T_{res}+DIFS+T_{B}&w.p.\quad  p_{b}
\end{matrix}\right.\label{eq:A},
\end{equation}
where $T_{res}$ is the residual lifetime of an ongoing packet transmission, and $T_{B}$ is the backoff duration.

Each slot in the backoff process can be interrupted by a transmission from other packets. During the interruption, the backoff counter is suspended. When the backoff counter is resumed, it starts from the beginning of the interrupted slot after deferring for a DIFS period. Therefore, the backoff duration $T_{B}$ is
\begin{equation}
T_{B}=\sum_{n=1}^{M}\left ( \sigma +T_I \right )\label{eq:B},
\end{equation}
where $\sigma$ is the duration of a time slot, r.v. $T_I$ is the interruption duration per slot, and r.v. $M$ is the backoff counter value. If no other vehicle send packets in a given slot, an interruption does not occur, which indicates $T_I$ is equal to zero. The slot will be interrupted when at least one another vehicle sends a packet in that slot. $T_I$ can be expressed as
\begin{equation}
T_I=\left\{\begin{matrix}
 0 &w.p. \quad \left ( 1-\rho\pi _{0} \right )^{N_{tr}-1} \\ 
 T_{tr}+ DIFS &w.p. \quad 1-\left ( 1-\rho\pi _{0} \right )^{N_{tr}-1}
\end{matrix}\right\}\label{eq:I}.
\end{equation}
Since $M$ an $T_I$ are two r.v.s, the backoff duration $T_B$ is sum of a random number of r.v.s. The mean of $T_B$ is found readily by using conditional expectation and as a result it is
\begin{equation}
\mathbb{E}\left [ T_{B} \right ]=\left ( \sigma +\mathbb{E}\left [ T_I \right ] \right )\mathbb{E}\left [ M \right ]\label{eq}.
\end{equation}
As $M$ is a r.v. which is uniformly distributed in the range $\left [ 0, CW-1 \right ]$, we can get
\begin{equation}
\mathbb{E}\left [ M \right ]=\frac{CW-1}{2}\label{eq:M}.
\end{equation}
Meanwhile, the mean of interruption time $T_I$ is
\begin{equation}
\mathbb{E}\left [ T_I \right ]=\left ( 1-\left ( 1-\rho\pi_{0} \right )^{N_{tr}-1} \right )(T_{tr}+DIFS)\label{eq:I}.
\end{equation}
From Eq. (\ref{eq:A}), the mean of access delay $T_A$ is obtained by
\begin{equation}
\mathbb{E}\left [ T_{A} \right ]=DIFS+p_b\left ( \mathbb{E}\left [ T_{B} \right ] +\mathbb{E}\left [ T_{res} \right ]\right )\label{eq},
\end{equation}
where $T_{res}$ follows the uniform distribution. Thus, the mean of $T_{res}$ is
\begin{equation}
\mathbb{E}\left [ T_{res} \right ]=\frac{T_{tr}}{2}+DIFS\label{eq}.
\end{equation}
Now we can get the mean delay $\mathbb{E}[T_s]$ which is actually equal to the average service time:
\begin{equation}
\mathbb{E}[T_s]=\mathbb{E}\left [ S \right ]=\mathbb{E}\left [ T_{A} \right ]+ T_{tr} \label{eq:E}.
\end{equation}

The reception delay $T_{re}$ describes how long other vehicles can receive a packet from the tagged vehicle, which includes the service time $S$ and the possible collision delay $T_{c}$. The collision delay is caused by the packet loss when collision occurs. Since the collision probability is $p_{c}$, the mean collision delay follows a geometric distribution and is given by
\begin{equation}
\mathbb{E}[T_{c}]=\frac{1}{\lambda}\sum_{n=1}^{\infty }np_{c}^{n}(1-p_{c})=\frac{p_{c}}{(1-p_{c})\lambda}\label{eq}.
\end{equation}
Thus, the mean reception delay is
\begin{equation}
\mathbb{E}[T_{re}]=\mathbb{E}[S]+\mathbb{E}[T_{c}].
\end{equation}

\section{IMPROVEMENT ON DSRC PERFORMANCE}
The objective of this section is to develop a distributed scheme - Semi-persistent Contention Density Control (SpCDC) aiming to improve the DSRC performance especially in heavy vehicle load scenarios. The tagged vehicle maintains a timeline and marks the slots when other vehicles generate their packets through the received packets in the previous periods. In a new transmission period, when the tagged vehicle receives packets from neighbor vehicles before it generates a packet, it will know the packets from those vehicles are no longer contending for channel access in this current period. The scenario of contending for channel access happens when the neighbor vehicles have generated packets but the tagged vehicle hasn't yet received them at the instant it generates a packet. By counting the number of these packets, the tagged vehicle will know the instantaneous contention density and determine its backoff counter\cite{gao2018contention}. 

\subsection{Analytical model for SpCDC scheme}
Denote the number of packets contending for channel access measured at the beginning of slot $k$ as $c\left ( k \right )$. Let $S\left ( k \right )=1$ and $S\left ( k \right )=0$ represent the events that slot $k$ is sensed busy and idle. If slot $k$ is sensed busy, the initial backoff counter of new generated packets arriving at slot $k\left [ m \right ]$ (There are $\frac{T_{tr}}{\sigma }$ mini-slots in slot $k, m\in V$, and $V=\left \{ 1,2,...,\frac{T_{tr}}{\sigma } \right \}$) will be stopped until the ongoing transmission ends. If slot $k$ is idle, $m=1$. The initial backoff counter $b\left ( k\left [ m \right ] \right )$ of a packet arriving at slot $k\left [ m \right ]$ depends on the instantaneous contention density. Denote the number of packets that arrives at slot $ k\left [ m \right ]$  measured at the m$th$ mini-slot in slot $k$ as $n_{a}\left ( k\left [ m \right ] \right )$. Denote the number of packets with their backoff counters reducing to 0 at slot $k$ as $n_{t}\left ( k\right )$. The framework of SpCDC is given in Algorithm 1
\begin{algorithm}
\caption{Framework of Semi-persistent Contention Density Control} 
\label{alg1}
\begin{algorithmic}
\REQUIRE Maintaining a list of timeline of packet generations based on the previous transmission periods.
\ENSURE A packet of tagged vehicle is generated and just arrives at the buffer, waiting to be sent.
\IF{$S(k) = 1$} 
\STATE $b\left ( k\left [ m \right ] \right )=C\cdot\left ( c\left ( k \right ) +\sum_{s=1}^{m}n_{a}\left ( k\left [ s \right ] \right )\right ), m\in V$ 
\IF {$\mathbb{R}=1$} 
\STATE $b(k[m])=b(k[m])+\omega, \ \omega \in \left \{-1, 0, 1\right \} $
\ELSE 
\STATE $b(k[m])=b(k[m])$
\ENDIF 
\STATE $S\left ( k+b\left ( k \left [ m \right ]\right ) \right )=1$
\STATE $c\left ( k+1 \right )=c\left ( k \right )+\sum _{s:s\epsilon V}n_{a}\left ( k\left [ s \right ] \right )-n_{t}\left ( k\right )$ 
\ELSE 
\STATE $b(k[1])=C \cdot (c(k)+n_{a}(k[1]))$
\IF {$\mathbb{R}=1$} 
\STATE $b(k[1])=b(k[1])+\omega, \ \omega \in \left \{-1, 0, 1\right \} $
\ELSE 
\STATE $b(k[1])=b(k[1])$
\ENDIF 
\STATE $S\left ( k+b\left ( k \left [ 1 \right ]\right ) \right )=1$
\STATE $c(k+1)=c(k)+n_{a}\left ( k\left [ 1 \right ] \right )$ 
\ENDIF 
\end{algorithmic}
\end{algorithm}
where $C$ is SpCDC protocol parameter, and $\mathbb{R}$ indicates the state whether the vehicle enters a new semi-persistent period. $\omega$ is the changed amount of the backoff counter value based on contention density at the beginning of each semi-persistent period. It is randomly selected from set $\left \{ -1, 0, 1 \right \}$ with equal probability. 

Since the expected change of $c\left ( k \right )$ in one slot is
\begin{equation}
\mathbb{E}\left \{ \Delta c\left ( k \right ) \right \}=\left\{\begin{matrix}
\lambda N_{tr}\sigma \quad if \ S\left ( k \right )=0\\ 
\lambda N_{tr}T_{tr}-n_{b} \quad if \ S\left ( k \right )=1
\end{matrix}\right\}\label{eq:Eck},
\end{equation}
where $\sigma$ is the duration of a time slot. $T_{tr}$ is the transmission delay. $n_{b}$ is the average number of packets in a busy slot. The probability of a slot being sensed idle and busy are given respectively by
\begin{equation}
P\left ( S\left ( k \right ) =0\right )=P_{ck\left ( 0 \right )}+\left ( 1-P_{ck\left ( 0 \right )} \right )\left ( 1-\gamma \right )\label{eq:Sk=0}
\end{equation}
\begin{equation}
P\left ( S\left ( k \right ) =1\right )=\left ( 1-P_{ck\left ( 0 \right )} \right )\gamma\label{eq:Sk=1},
\end{equation}
where $P_{ck\left ( 0 \right )}$ is the probability of no packet contending for channel access at slot k, i.e., $c\left ( k \right )=0$. $\gamma$ indicates how many packets each backoff slot accommodates in average, and it is an approximate probability that the slot $k$ is sensed busy given at least one contending packet. Since the expected change of $c\left ( k \right )$ should be equal to 0 in the steady state, it holds
\begin{equation}
\begin{aligned}
&\mathbb{E}\left \{ \Delta c\left ( k \right )|S(k)=0\right \}P\left ( S\left ( k \right ) =0\right )\\& + \mathbb{E}\left \{ \Delta c\left ( k \right ) |S(k)=1\right \}P\left ( S\left ( k \right ) =1\right )=0\label{eq:sumE}.
\end{aligned}
\end{equation}
Therefore, $\gamma$ can be obtained by plugging Eq. (\ref{eq:Eck}), (\ref{eq:Sk=0}) and (\ref{eq:Sk=1}) into Eq. (\ref{eq:sumE})
\begin{equation}
\gamma=\frac{\lambda N_{tr}\sigma }{\left ( 1-P_{ck\left ( 0 \right )} \right )\left ( n_{b} -\lambda N_{tr}\left ( T_{tr}-\sigma  \right )\right )}\label{eq}.
\end{equation}

The average number of packets in a busy slot $n_{b}$ is greater than 1 due to packet collision. Assume each collision only involves two packets with collision probability $P_{c}$, $n_{b}$ is given by
\begin{equation}
n_{b}=1+P_{c}\label{eq}.
\end{equation}

Now we start to derive the mean delay for channel access which includes the busy slots and idle slots during the backoff process. Suppose the arrival of a new packet is uniformly distributed in a busy slot, the mean of duration of busy slots $T_{db}$ is
\begin{equation}
\begin{aligned}
&\mathbb{E}[T_{db}]=\left(1+\sum_{j=1}^{N_{tr}-1}P_{ck\left ( j \right )}\left ( j-\frac{1}{2} \right )\right)T_{tr}\\&=\left ( c_{s} +\frac{1}{2}\left ( 1+P_{ck\left ( 0 \right )} \right )\right )T_{tr}\label{eq:T_{db}},
\end{aligned}
\end{equation}
where $c_{s}$ is the mean contention density, and $P_{ck\left ( j \right )}=P\left ( c\left ( k\right ) =j\right )$, i.e., the probability of $j$ packets contending for channel access. According to the computed initial backoff counter and the number of busy slots, the mean of duration of idle slots $T_{di}$ is
\begin{equation}
\mathbb{E}[T_{di}]=\left ( C \cdot (c_{s}+1) -c_{s}\right )\sigma \label{eq:T_{di}}.
\end{equation}
Given $\mathbb{E}[T_{db}]$ and $\mathbb{E}[T_{di}]$, we can get the mean delay $\mathbb{E}[T_{d}]$
\begin{equation}
\mathbb{E}[T_{d}]=\mathbb{E}[T_{db}]+\mathbb{E}[T_{di}]\label{eq:T_{d}},
\end{equation}
while the mean reception delay is
\begin{equation}
\mathbb{E}[T_{re}]=\mathbb{E}[T_{d}]+\mathbb{E}[T_{c}]\label{eq}.
\end{equation}

Since each packet arrives every $1/\lambda$ seconds, the probability of a packet staying at the buffer is $\frac{\mathbb{E}[T_{d}]}{1/\lambda}$. The mean contention density should satisfy
\begin{equation}
c_{s}=(N_{tr}-1)\frac{\mathbb{E}[T_{d}]}{1/\lambda}=\lambda (N_{tr}-1)\mathbb{E}[T_{d}]\label{eq}.
\end{equation}
We can also obtain the mean contention density $c_{s}^{'}$ in IEEE 802.11p based on Eq. (\ref{eq:M}) and (\ref{eq:I}):
\begin{equation}
c_{s}^{'}=\frac{(CW-1)\left ( 1-\left ( 1-\rho\tau \right )^{N_{tr}-1} \right )}{2}\label{eq}.
\end{equation}
Given the probability of one packet staying at the buffer, the probability that no packet is contending for channel access over $N_{tr}-1$ vehicles is
\begin{equation}
P_{ck\left ( 0 \right )}=\left ( 1-\lambda \mathbb{E}[T_{d}] \right )^{N_{tr}-1}=\left ( 1-\frac{c_{s}}{N_{tr}-1} \right )^{N_{tr}-1}\label{eq}.
\end{equation}

For an arbitrary $k$, we consider the worst case so that we can derive the upper bound of collision probability. In the worst case, the initial backoff counter of an incoming packet always holds $b(k[m])< CW(k)$ where $CW(k)$ is the contention window at slot $k$. If a collision occurs in the initial slot in the backoff process, the collision probability will be $\gamma$. If no collision occurs in the initial backoff slot, the collision may occur in the remaining $C\left ( c_{s} +1\right )-1$ slots. Suppose slots are independent with each other, the collision probability in each slot is given by $ 1-\left ( 1-\gamma \right ) ^{c_{s}}$, Thus, the upper bound of collision probability is
\begin{equation}
\begin{aligned}
P_{c}^{upper}=(1-P_{ck(0)})(\gamma+(1-\gamma) ( 1-\left ( 1-\gamma \right ) ^{c_{s}})^{C(c_{s}+1))-1})\label{eq},
\end{aligned}
\end{equation}
and the lower bound of PDR is
\begin{equation}
PDR^{lower}=1-P_{c}^{upper}\label{eq}.
\end{equation}

%% file: results.tex
\section{RESULTS OF ANALYTICAL MODEL AND SIMULATION}

In this section, we present a simulation setup used to validate our analytical model and give validation results. The computation for analytic models with corresponding simulations are conducted in Matlab. All assumptions are the same in the simulation and analytical models. Each vehicle on the lanes is equipped with DSRC wireless capability with perfect PHY-layer performance. Since vehicle can communicate with each other in a fully connected network, the location of each vehicle doesn't impact their performance. We also use Table I's parameters for the simulation in the SpCDC scheme, where the protocol parameter $C$ = 3 and a complete semi-persistent period is 1 second for each vehicle. 

\begin{table}[t]
\captionsetup{font={scriptsize}}
\caption{DSRC communication parameters}
\begin{center}
\begin{tabular}{|c|c|}
\hline
\textbf{Parameters}&\textbf{Values} \\
\hline
Packet length (payload), $E[P]$&200, 400 bytes\\
\hline
PHY preamble &28 us\\
\hline
MAC header & 50 bytes\\
\hline
Packet transmission frequency, $\lambda$& 2, 10 pps\\
\hline
Slot time, $\sigma$& 16 us\\
\hline
Propagation delay, $\delta $& 0 us\\
\hline
PLCP header& 4 us\\
\hline
Contention window, $CW$& 16\\
\hline
Number of vehicles & 10, 20, ..., 200 \\
\hline
DIFS& 64 us\\
\hline
Data rate, $R_{d}$&6, 12, 24 Mbps \\
\hline
\end{tabular}
\label{tab1}
\end{center}
\end{table}

\begin{figure}[h] 
  \centering
  \subfigure[Mean delay]{\includegraphics[width=0.75\linewidth]{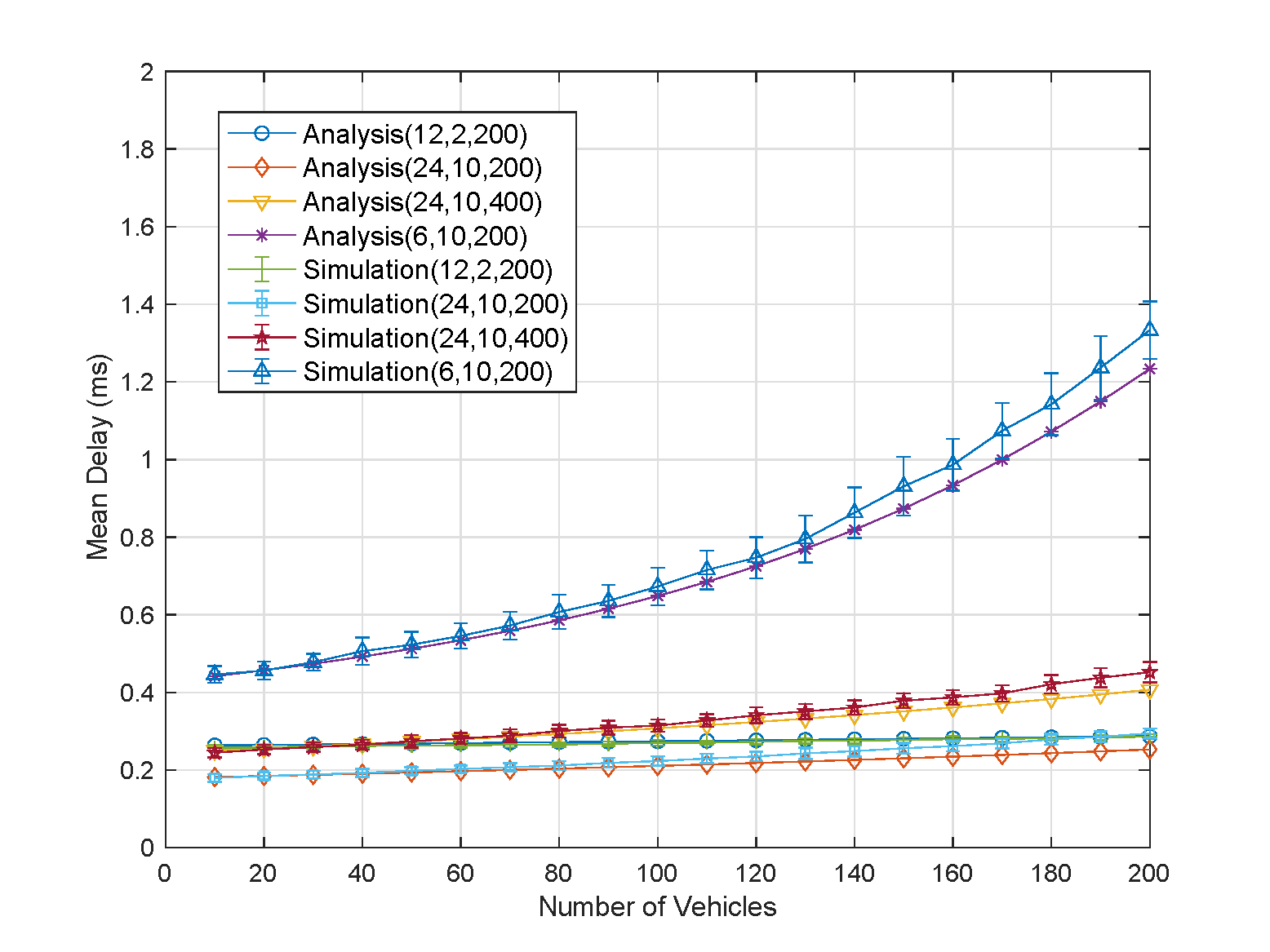}}\quad
  \subfigure[PDR]{\includegraphics[width=0.75\linewidth]{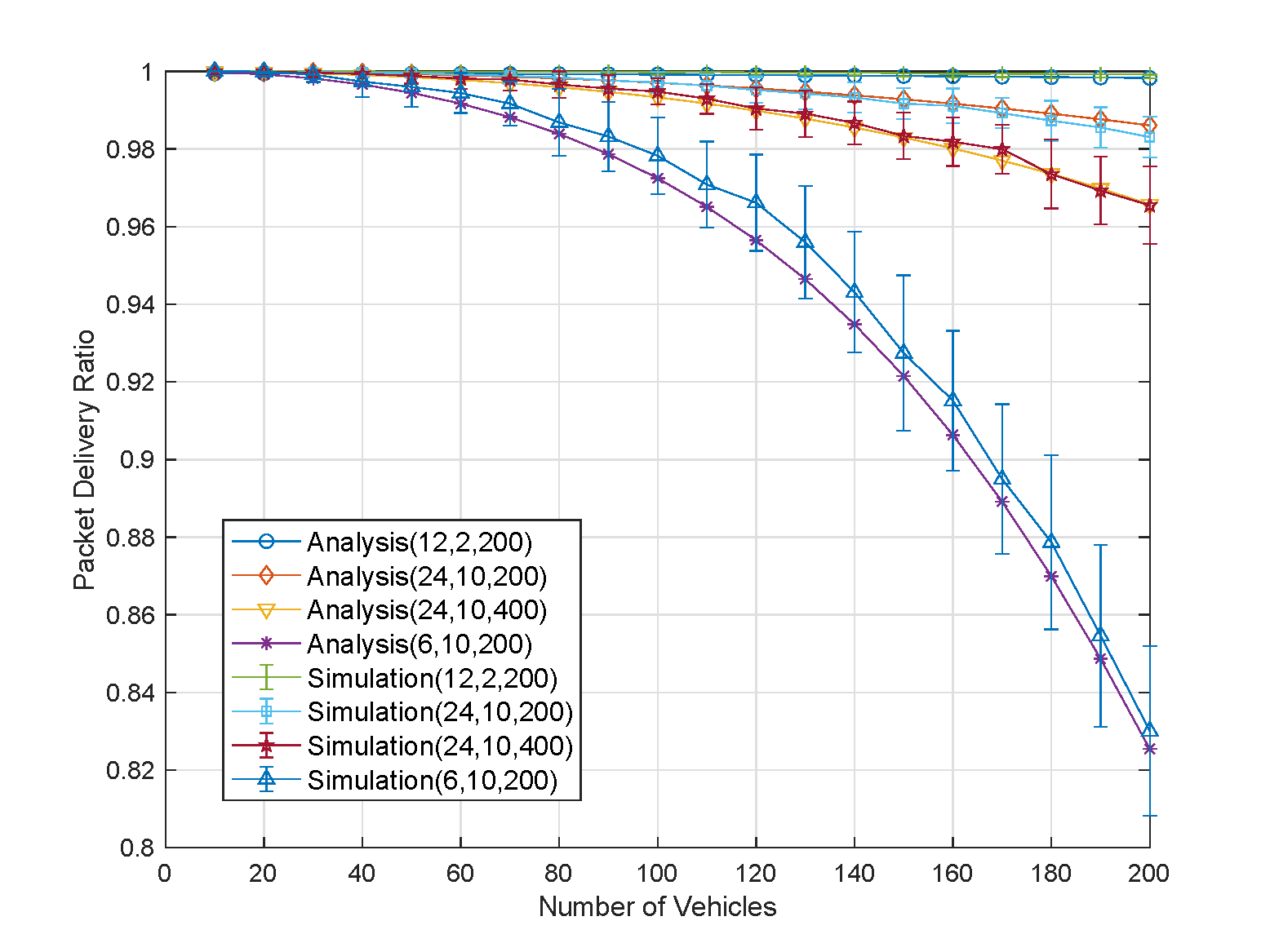}}
  \caption{Performance in IEEE 802.11p}
  \label{fig:Per}
\end{figure}

Fig.\ref{fig:Per} shows the DSRC performance in IEEE 802.11p as a function of number of vehicles, with different curves parameterized by data rate $R_{d}$ (in megabits per second), packet transmission frequency $\lambda$ (packets per second) and mean packet length $E[P]$ (in bytes). The analytic model agrees well with the simulation results. In the plotted range, the average delay increases almost linearly with the vehicle density except for the case of 6 Mbps/10 packets per second/ 200 bytes. Meanwhile, the PDR in this case also drops markedly with the increasing vehicle load. The reason why this case differs from other cases is caused by more interruptions during backoff process and higher transmission delay.
\begin{figure}[h]
\centering
\includegraphics[width=0.75\linewidth]{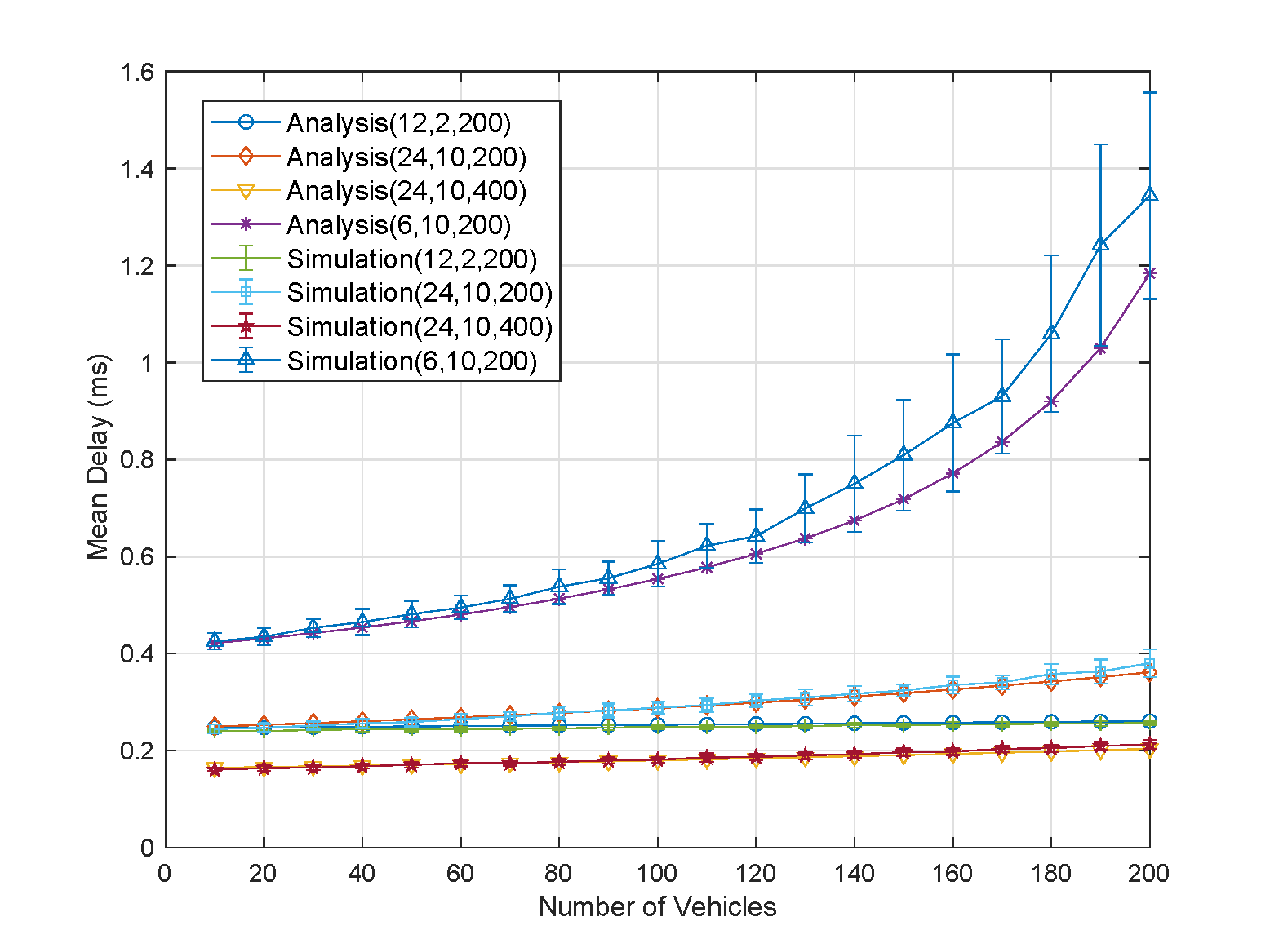}
\caption{Mean delay in SpCDC}
\label{fig:spic_delay}
\end{figure}

We observe the improvement on DSRC performance according to the developed model. First, as Fig.\ref{fig:spic_delay} shows, the analytical model matches well with the simulation results of the mean delay, which validates our model. Next, we only focus on a typical case (6 Mbps/10 pps/200 bytes), which best approximates the parameters in a real situation.
\begin{figure}[htbp] 
  \centering
  \subfigure[Mean delay]{\includegraphics[width=0.75\linewidth]{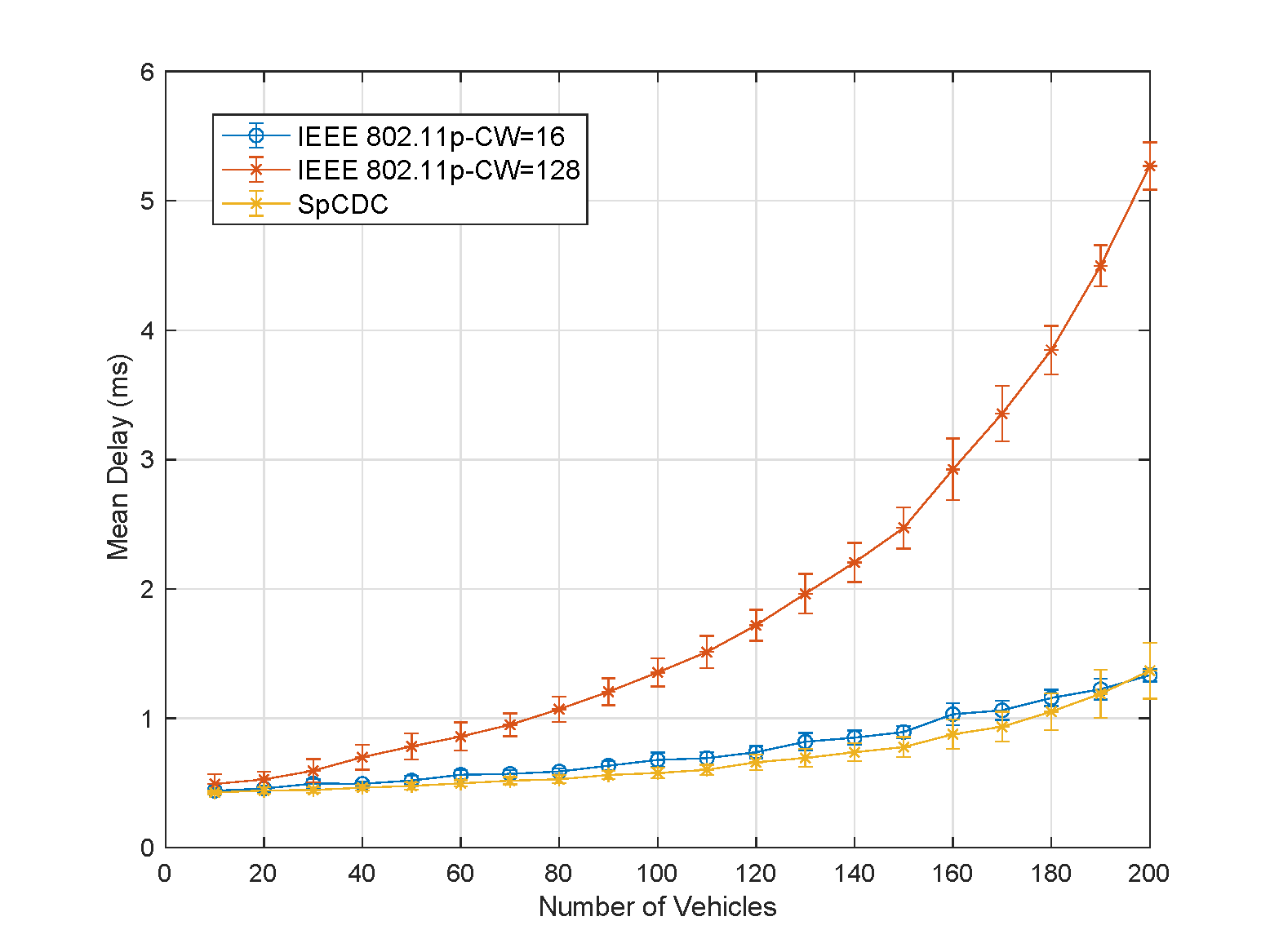}}\quad
  \subfigure[Contention density]{\includegraphics[width=0.75\linewidth]{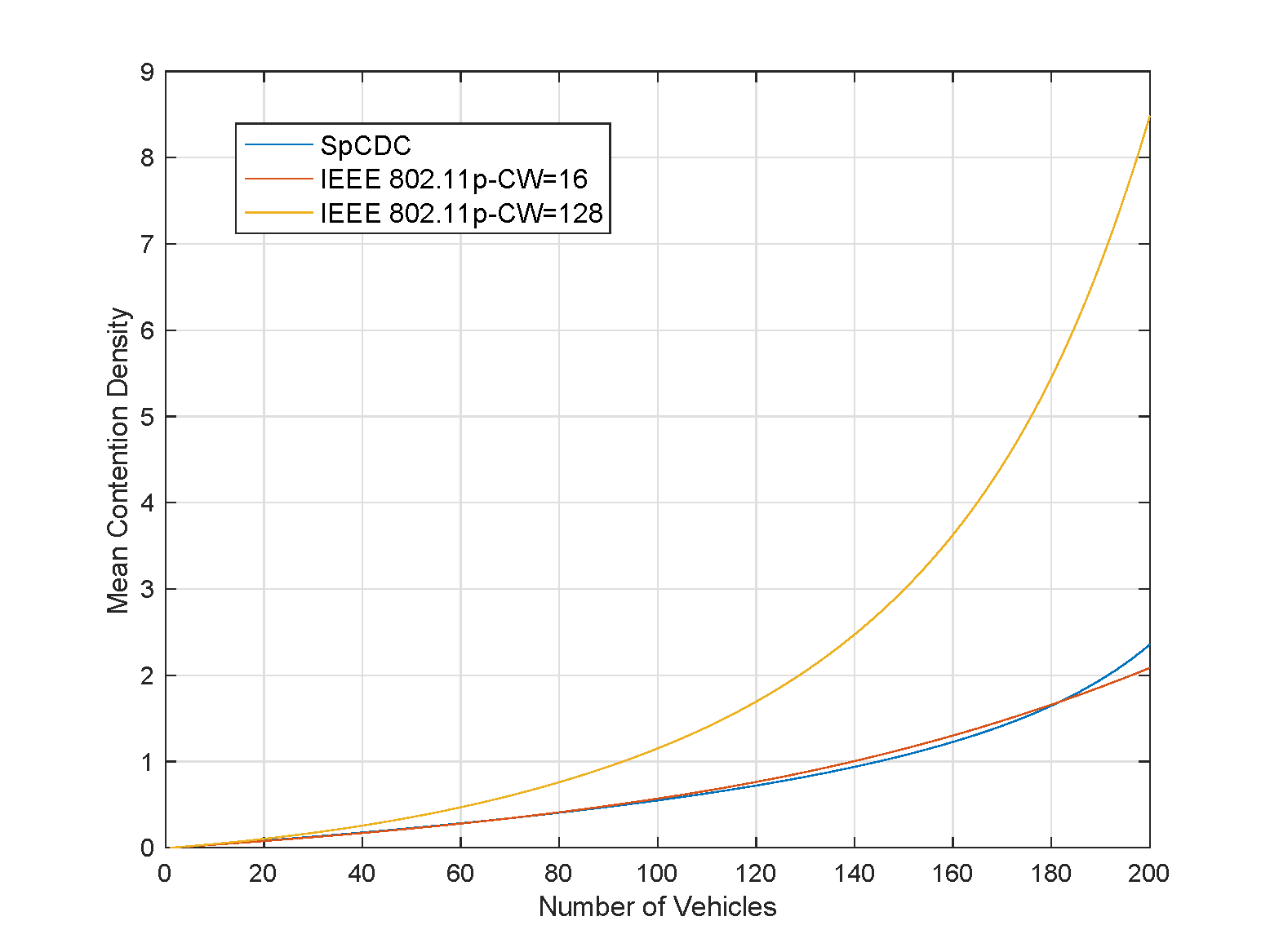}}
  \caption{Comparison of the contention density }
  \label{fig:comp_delay}
\end{figure}
Fig.\ref{fig:comp_delay}(a) shows the mean delay in IEEE in 802.11p with different contention windows and SpCDC scheme. The mean delay in SpCDC scheme is always below that in 802.11p with CW=128 while being very close to that in 802.11p with CW=16. Fig.\ref{fig:comp_delay}(b) shows the contention density among them. When the number of vehicles is 200, the contention density in SpCDC scheme is around seven fewer than that in IEEE 802.11p with CW=128. Since the transmission delay is around 0.5 ms, the mean delay difference between 802.11p with CW=128 and SpCDC scheme will be more than 3 ms.

\begin{figure}[htbp] 
  \centering
  \subfigure[PDR]{\includegraphics[width=0.75\linewidth]{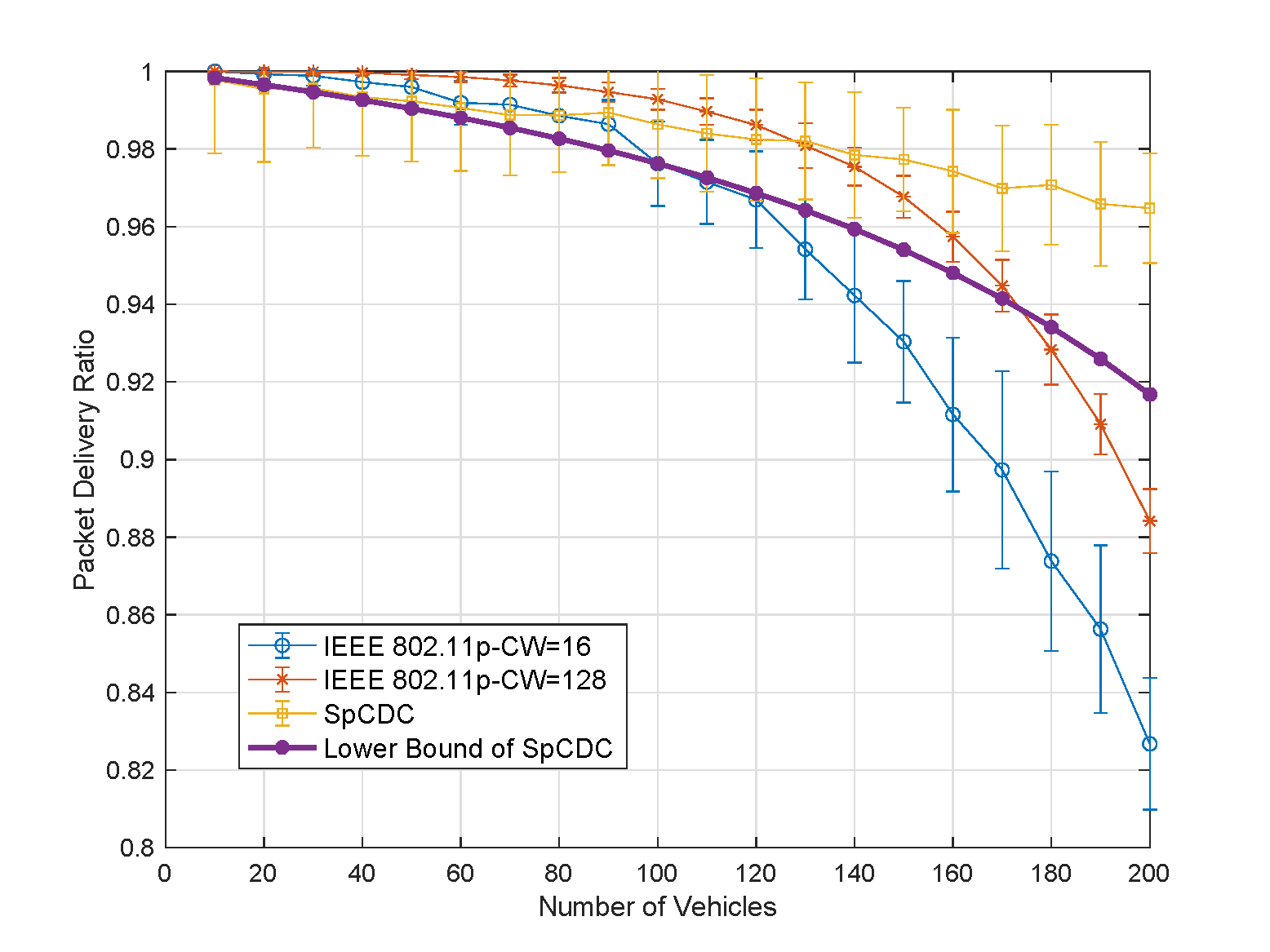}}\quad
  \subfigure[Mean reception delay]{\includegraphics[width=0.75\linewidth]{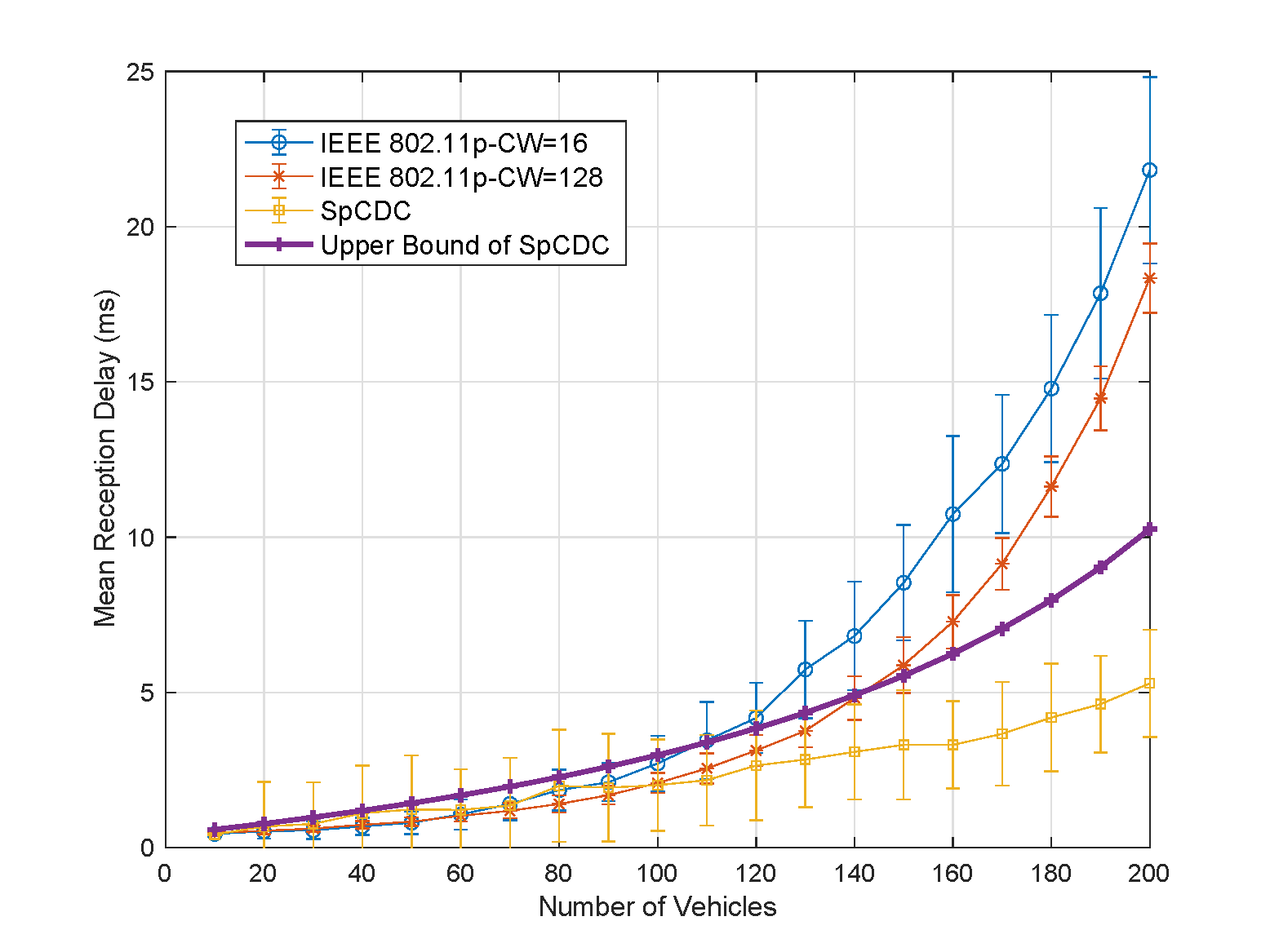}}
  \caption{Comparison of the PDR and reception delay}
  \label{fig:Comparison}
\end{figure}

Fig.\ref{fig:Comparison}(a) presents the simulation results for PDR in IEEE 802.11p and SpCDC scheme and the analytical lower bound of PDR for SpCDC scheme. The PDR in SpCDC increases nearly 15\% compared with that in IEEE 802.11p with CW=16 and 10\% with CW=128 in heavy vehicle loads. Besides, we can also observe the analytical lower bound is not very tight especially in heavy vehicle loads since the lower bound of PDR is derived under the assumption of worst case. Nevertheless, even the lower bound lies above the performance of PDR in IEEE 802.11p.
Fig.\ref{fig:Comparison}(b) provides the simulation result for the mean reception delay between SpCDC and IEEE 802.11p. As the result shows, the mean reception delay in SpCDC scheme is much lower than IEEE 802.11p with different contention windows, even the upper bound of mean delay in SpCDC is lower nearly 50\% than IEEE 802.11p with CW=128. This result indicates DSRC adopting SpCDC scheme can receive more timely BSMs in a long period compared with IEEE 802.11p. In other words, SpCDC scheme provides more reliable road safety than IEEE 802.11p by lowing down the mean reception delay for each vehicle.

%% file: conclusion.tex
\section{CONCLUSION}
In this paper, we first focused on the performance analysis of DSRC performance adopting IEEE 802.11p in periodic broadcast mode. With the assumption of a perfect PHY performance and the fixed point method, we presented the PDR and packet delay in a fully connected network. Our analytic model provided a good match with simulation results. Then we developed the SpCDC scheme to improve DSRC performance. By comparing the SpCDC scheme with IEEE 802.11p with some metrics such as PDR and mean reception delay, we can verify that SpCDC improves DSRC performance. Furthermore, it is possible to partially adjust this scheme which can be applied in the Listen Before Talk protocol based short-term sensing in NR V2X.

\section*{ACKNOWLEDGEMENTS}
The authors would like to express the special thanks to Prof. Randall Berry and Prof. James Ritcey for helping discuss and review this work!